\documentclass[12pt, draftclsnofoot, onecolumn]{IEEEtran}
\usepackage{amsmath}
\usepackage{amssymb}
\usepackage{amsthm}
\usepackage{cite}
\usepackage{color}
\usepackage{xcolor}
\usepackage{caption}
\usepackage{epsfig,latexsym}
\usepackage{float}
\usepackage{fancyhdr}
\usepackage{graphicx}
\usepackage{indentfirst}
\usepackage{mdwmath}
\usepackage{mdwtab}
\usepackage{subfigure}
\usepackage{setspace}
\usepackage{times}
\usepackage{url}
\usepackage{multirow}

\newtheorem{remark}{Remark}
\newtheorem{theorem}{Theorem}

\newtheorem{lemma}{Lemma}

\newtheorem{corollary}{Corollary}

\newtheorem{proposition}{Proposition}

\begin{document}

\title{MIMO-NOMA Networks Relying on Reconfigurable Intelligent Surface: A Signal Cancellation Based Design}
\author{Tianwei Hou,~\IEEEmembership{Student Member,~IEEE,}
        Yuanwei Liu,~\IEEEmembership{Senior Member,~IEEE,}
        Zhengyu Song,
        Xin Sun,
        and Yue Chen,~\IEEEmembership{Senior Member,~IEEE,}

\thanks{This work is supported by the National Natural Science Foundation of China under Grant 61901027. Part of this paper was submitted at the IEEE Global Communication Conference, Taipei, Taiwan, China, Dec. 2020~\cite{Hou_RIS_SCB_GLOBECOM}.}
\thanks{T. Hou, Z. Song and X. Sun are with the School of Electronic and Information Engineering, Beijing Jiaotong University, Beijing 100044, China (email: 16111019@bjtu.edu.cn, songzy@bjtu.edu.cn, xsun@bjtu.edu.cn).}
\thanks{Y. Liu and Yue Chen are with School of Electronic Engineering and Computer Science, Queen Mary University of London, London E1 4NS, U.K. (e-mail: yuanwei.liu@qmul.ac.uk, yue.chen@qmul.ac.uk).}
}

\maketitle

\begin{abstract}
Reconfigurable intelligent surface (RIS) technique stands as a promising signal enhancement or signal cancellation technique for next generation networks. We design a novel passive beamforming weight at RISs in a multiple-input multiple-output (MIMO) non-orthogonal multiple access (NOMA) network for simultaneously serving paired users, where a signal cancellation based (SCB) design is employed. In order to implement the proposed SCB design, we first evaluate the minimal required number of RISs in both the diffuse scattering and anomalous reflector scenarios. Then, new channel statistics are derived for characterizing the effective channel gains. In order to evaluate the network's performance, we derive the closed-form expressions both for the outage probability (OP) and for the ergodic rate (ER). The diversity orders as well as the high-signal-to-noise (SNR) slopes are derived for engineering insights. The network's performance of a finite resolution design has been evaluated. Our analytical results demonstrate that: i) the inter-cluster interference can be eliminated with the aid of large number of RIS elements; ii) the line-of-sight of the BS-RIS and RIS-user links are required for the diffuse scattering scenario, whereas the LoS links are not compulsory for the anomalous reflector scenario. 
\end{abstract}


\section{Introduction}

The demanding for having high spectrum efficiency (SE) and energy efficiency (EE) has been rapidly increasing in next-generation (NG) networks~\cite{5G_NR}. A promising access technique, non-orthogonal multiple access (NOMA), has been proposed, which is capable of simultaneously serving multiple users at different quality-of-service requirements~\cite{NOMA_mag_Ding,PairingDING2016,Massive_NOMA_Cellular_IoT}. Multiple users share the same time/frequency/code resource block by with the aid of superposition coding (SC) at the transmitter and successive interference cancellation (SIC) at the receiver by capitalizing the difference of users' channel state information (CSI)~\cite{NOMA_5G_beyond_Liu,Islam_NOMA_survey,NOMA_large_heter}.

Since multiple antennas (MAs) offer extra diversity by its spacial domain, MA techniques are of significant importance. The application of MA techniques assisted NOMA networks has attached significant attention. In classic multiple-input multiple-output (MIMO) designs, the base (BS) equipped with $M$ transmitting antennas (TAs) is capable of transmitting maximal $M$ interference-free beams. However, since multiple users are paired to perform NOMA in each cluster, the BS equipped with $M$ TAs has to compulsorily serve $KM$ users simultaneously, where $K$ denotes the number of users in each cluster. Hence, how to design interference-free beamforming becomes an interesting problem, which is valuable to examine. A zero-forcing-based (ZF-based) design was proposed in~\cite{zero-forcing_detection}, where the active beamforming at the BS is an identity matrix. However, there are two drawbacks, where 1) the number of receiving antennas (RAs) has to be higher than that of the BS to ensure the existence of a solution; and 2) due to the property of singular value decomposition, the antenna gain of the ZF-based design can be obtained as $L-M+1$, where $L$ denotes the number of RAs. Then a signal-alignment-based (SA-based) design is proposed to release the constraint of the number of RAs~\cite{signal-alignment-design}, which relies on the gain shifting of user's channel matrixes. By doing so, the small-scale channel gains of $K$ users in each cluster are identical, but correspond to different distances, and hence $MK$ users can be treated as $M$ users. It is demonstrated that multi-user NOMA networks may be not practical due to high computational complexity~\cite{Capacity_Zeng}. Numerous applications related to MIMO-NOMA were proposed, i.e. MIMO-NOMA enhanced physical layer security networks~\cite{Liu_physical_scurity_NOMA} and MIMO-NOMA enhanced simultaneous wireless information and power transfer (SWIPT) networks~\cite{Liu_Coop_NOMA_SWIPT}. However, as mentioned above, there were many constrains on the number of RAs in the previous designs.

Recently, reconfigurable intelligent surface (RIS) technique stands as the next generation relay technique, also namely relay 2.0, received considerable attention due to its high EE~\cite{LIS_zhangjiayi_mag,LIS_smart,RIS_mag_basar,5G_NR_2}. The RIS elements are capable of independently shifting the signal phase and absorbing the signal energy, and hence the reflected signals can be boosted or diminished for wireless transmission~\cite{LIS_magazine_multi_scenarios,reconfig_meta_surf_1,reconfig_meta_surf_2}.
By doing so, numerous application scenarios have been considered, e.g. RIS-aided coverage enhancement. The RIS elements are normally deployed on the building or on the wall~\cite{Hou_RIS_MIMO_global_algrithm}. A novel three-dimension design for aerial RIS network was proposed in~\cite{ZhangRui_aerial_RIS}, where RIS elements are employed at aerial platforms, and hence a full-angle reflection can be implemented. Currently, RIS networks are simply separated into two categories~\cite{Renzo_Mirror_or_Scatterer}, i.e. anomalous reflector or diffuse scatterer for mmWave and sub-6G networks, respectively. The coverage distance is reduced in mmWave networks~\cite{Renzo_mmwave_signal_enhancement}, and hence more users are located in coverage-holes compared to conventional networks. Thus, reflected signals can be aligned by RISs for serving users located in the coverage-holes~\cite{Lv_coverage_enhancement}.



NOMA and RIS techniques can be naturally integrated for enhancing both SE and EE. The RISs can be deployed for the cell-edge users in the NOMA networks, where the reflected signal cannot be received at the cell-center users~\cite{DING_RIS_NOMA_letter}. An one-bit coding scheme was invoked in the RIS-aided NOMA networks, where imperfect SIC scenario was evaluated in~\cite{Xinwei_NOMA_1bit_coding}. Since both the BS and RISs are pre-deployed, and hence the line-of-sight (LoS) links between the BS and RISs are expected for improving desired signal power~\cite{MISO_with_directlink}. The Rician fading channels were utilized for illustrating the channel gain of both the BS-RIS and RIS-user links in~\cite{RIS_NOMA_Rice}. A SISO-NOMA network was proposed in~\cite{Hou_RIS_MIMO_NOMA_prioritized_algrithm}, where a prioritized design was proposed for further enhancing the network's SE and EE. However, previous contributions mainly focus on the signal enhancement based (SEB) designs, where signals are boosted at the user side or at the BS side.

\subsection{Motivations and Contributions}

Previous contributions mainly focus on the SEB designs, whilst there is a paucity of investigations on the signal-cancellation-based (SCB) design of the RIS-aided networks. Inspired by the concepts of the signal cancellation~\cite{Renzo_PHY_security_confe,PLS_LIS_ZhangRui}, we propose a novel SCB design concept, which provides the desired degree of flexibility for the RIS-aided networks. In the MIMO-NOMA networks, one of key challenge is to eliminate the inter-cluster interference. Hence, in order to illustrate the potential benefits provided by RISs, a RIS-aided SCB design in MIMO-NOMA networks is proposed for comprehensively analyzing the performance of the networks. 

Against to above background, our contributions can be summarized as follows:
\begin{itemize}
  \item We propose a novel SCB design in RIS-aided MIMO-NOMA networks, where the inter-cluster interference can be eliminated for enhancing the network's performance. The impacts of both the diffuse scattering and anomalous reflector scenarios are exploited. The impact of the proposed design on the attainable performance is characterized.
  \item We first derive the minimal required number of RISs for implementing the proposed SCB design. For the ideal-RIS (I-RIS) cases, our analytical results illustrate that the inter-cluster interference can be beneficially eliminated. We then evaluate the impact of non-ideal-RIS (NI-RIS) cases by finite resolution techniques.
  \item Explicitly, we derive closed-form expressions of both the OP and of the ER for the proposed SCB design. The exact closed-form expressions of the OP and of the ER are derived in both the diffuse scattering and anomalous reflector scenarios. The diversity orders and high-SNR slopes are derived based on the OP and ER. The results confirm that the diversity order is obtained as the number of RAs.
  \item Our analytical results illustrate that the proposed SCB design is capable of 1) releasing the constraint of the number of RAs; 2) perfectly eliminating the inter-cluster interference in the I-RIS cases. The simulation results confirm our analysis, illustrating that: 1) the outage floors and ergodic ceilings occur for the NI-RIS cases; 2) the proposed SCB design does not rely on a high number of RISs for increasing both the OP and ER. 3) the proposed SCB design is capable of outperforming the classic ZF-based and SA-based designs.
\end{itemize}
\subsection{Organization and Notations}

In Section \uppercase\expandafter{\romannumeral2}, a SCB design is investigated in RIS-aided MIMO-NOMA networks. The minimal required number of RISs is analyzed. Then the passive beamforming weight at RISs is proposed for implementing SCB design. Analytical results are presented in Section \uppercase\expandafter{\romannumeral3} to show the performance of the proposed SCB design. In Section \uppercase\expandafter{\romannumeral4}, the numerical results are provided for verifying our analysis. Then Section \uppercase\expandafter{\romannumeral5} concludes this article. 
$\mathbb{P}(\cdot)$ denotes the probability, and $\mathbb{E}(\cdot)$ denotes the expectation. 
${{\bf{H}}^{{T}}}$, ${{\bf{H}}^{{H}}}$, $rank({\bf{H}})$ denote the transpose and conjugate transpose, as well as rank of the matrix $\rm \bf H$.
$\left\| {\bullet } \right\|_2^2$ denotes the Frobenius norm. Table~\ref{TABLE OF NOTATINS} lists some critical notations used in this article.

\begin{table}
\caption{\\ TABLE OF NOTATIONS}
\centering
\begin{tabular}{|l|r|}
\hline
$R_{m,k}$ & The target rate of user $k$ in cluster $m$.\\
\hline
$\mathcal{K} _1$ &Fading factor between the BS and RISs.\\
\hline
$\mathcal{K}_2$ &Fading factor between the RISs and users.\\
\hline
$M$ & The number of TAs.\\
\hline
$L$ & The number of RAs.\\
\hline
$K$ & The number of users in each cluster.\\
\hline
$N$ & The number of RISs.\\
\hline
$\rm \bf H$  & The channel matrix of the BS-RIS link.\\
\hline
${\rm \bf G}_{m,k}$ & The channel matrix between the RISs and user $k$ in cluster $m$.\\
\hline
${\rm \bf W}_{m,k}$ & The channel matrix between the BSs and user $k$ in cluster $m$.\\
\hline
$\rm \bf \Phi$ & The effective matrix of RISs.\\
\hline
\end{tabular}
\label{TABLE OF NOTATINS}
\end{table}

\section{System Model}

\begin{figure}[t!]
\centering
\includegraphics[width =4 in]{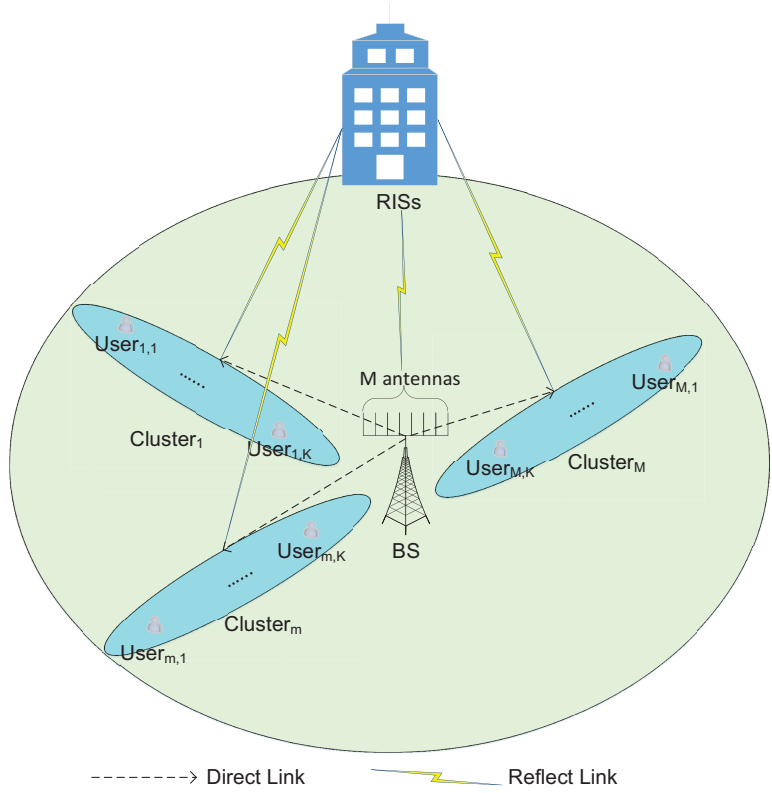}
\caption{System model of the RIS-aided MIMO-NOMA networks.}
\label{system_model}
\end{figure}

Let us focus our attention on a MIMO-NOMA network in downlink communication, where a BS equipped with $M$ TAs is simultaneously communicating with $MK$ users each equipped with $L$ RAs by utilizing the power-domain NOMA techniques. For simplicity, $MK$ users are separated into $M$ clusters, and $K$ users are paired in each cluster. We have $N$ intelligent surfaces at a proper location with $N \ge 1$ for implementing wireless communication. By properly controlling the  amplitude coefficients and phase shifts of each RIS element, the signals can be beneficially manipulated. The system model is illustrated in Fig.~\ref{system_model}.

\subsection{System Description of RIS-aided MIMO-NOMA Networks}

Since the LoS link does not exist in the BS-user link, the small-scale fading matrix between the BS and user $k$ in cluster $m$ is defined by Rayleigh fading channels, which can be expressed as
\begin{equation}\label{channel matrix direct link}
{{\rm \bf{W}}_{m,k}} = \left[ {\begin{array}{*{20}{c}}
{w_{1,1}^{m,k}}& \cdots &{w_{1,M}^{m,k}}\\
 \vdots & \cdots & \vdots \\
{w_{L,1}^{m,k}}& \cdots &{w_{1,1}^{m,k}}
\end{array}} \right],
\end{equation}
where ${{\rm \bf W}_{m,k}} $ is a $(L \times M)$ element matrix. Note that the probability density function (PDF) of the Rayleigh fading channel gains is given by
\begin{equation}\label{channel matrix,Rayleigh}
{f}(x) = {e^{ - {{x}}}}.
\end{equation}
The large-scale fading between the BS and user $k$ in cluster $m$ is given by
\begin{equation}\label{large scale fading between BS and user}
{L_{{\rm {b}},m,k}} = d_{ {\rm {b}},m,k}^{ - \alpha_3 },
\end{equation}
where $\alpha_3$ denotes the path loss exponent between the BS and user $k$ in cluster $m$.

Since the LoS links are expected for  both the BS-RIS and RIS-user links, the channel matrixes are defined by Rician fading channels as follows:
\begin{equation}\label{channel matrix,BS to LIS}
{\rm \bf{H}} = \left[ {\begin{array}{*{20}{c}}
{{h_{1,1}}}& \cdots &{{h_{1,M}}}\\
 \vdots & \cdots & \vdots \\
{{h_{N,1}}}& \cdots &{{h_{N,M}}}
\end{array}} \right],
\end{equation}
where ${\rm \bf H}$ is a $(N \times M)$ Rician fading channel gains, which can be modeled as follows:
\begin{equation}\label{Rice channel gain}
h_{n,m} = \sqrt {\frac{ \mathcal{K}_1 }{{\mathcal{K}_1 + 1}}} h_{n,m}^{{\rm{LoS}}} + \sqrt {\frac{1}{{\mathcal{K}_1 + 1}}} h_{n,m}^{{\rm{NLoS}}},
\end{equation}
where $\mathcal{K}_1$ denotes the Rician factor of the BS-RIS link. $h_{n,m}^{{\rm{LoS}}} $ and $h_{n,m}^{{\rm{NLoS}}} $ denote the LoS and Non-LoS (NLoS) components, respectively.

%

Similar to~\eqref{channel matrix,BS to LIS}, the small-scale fading matrix between the RISs and user $k$ in cluster $m$ is defined as
\begin{equation}\label{channel matrix,LIS to w user}
{{\bf{G}}_{m,k}} = \left[ {\begin{array}{*{20}{c}}
{g_{1,1}^{m,k}}& \cdots &{g_{1,N}^{m,k}}\\
 \vdots & \cdots & \vdots \\
{g_{L,1}^{m,k}}& \cdots &{g_{L,N}^{m,k}}
\end{array}} \right],
\end{equation}
where ${{\rm \bf{G}}_{m,k}}$ is $(L \times N)$ matrix whose elements represent Rician fading channel gains with fading parameter $\mathcal{K}_2$.
Similarly, the channel gains of the RIS-user link is given by
\begin{equation}\label{Rice channel gain}
g_{l,n}^{m,k} = \sqrt {\frac{ \mathcal{K}_2 }{{\mathcal{K}_2 + 1}}} g_{l,n}^{m,k,{\rm{LoS}}} + \sqrt {\frac{1}{{\mathcal{K}_2 + 1}}} g_{l,n}^{m,k,{\rm{NLoS}}},
\end{equation}
where $\mathcal{K}_2$ denotes the Rician factor of the BS-RIS link. $g_{l,n}^{m,k,{\rm{LoS}}} $ and $g_{l,n}^{m,k,{\rm{NLoS}}} $ denote the LoS and NLoS components, respectively.


In this article, $d_1$ and $d_{m,k}$ represent the distances of the BS-RIS and RIS-user links. In order to limit the inter-cluster interference, the received signal power levels of both the BS-user and reflected links are expected to the same order.
Recently, anomalous reflecting and diffuse scattering scenarios are presented for the mmWave and sub-6G networks, respectively~\cite{renzo_RIS_relay}. On the one hand, the size of RISs is comparable with the wavelength for the sub-6G networks, and hence the RISs are expected to be diffusers. Such as radar networks, the path loss of reflected links are expected as product-distance law~\cite{RIS_diffuse_model_book}. On the other hand, the wavelength is sufficiently small compared with the size of RISs, hence the theory of geometric optics is capable of modeling the path loss, where the sum-distance law of a specular reflection holds in the anomalous reflector scenarios based on the generalized Snell's law. We then study the feasibility of two scenarios, where the RISs act as anomalous reflectors or diffuse scatterers.

1) \emph{Diffuse Scattering Scenario:} When the size of RISs is comparable with the wavelength, where the size of RIS elements is usually set to $\frac{1}{4}$ wavelength, the RISs are considered as diffuse scatters~\cite{renzo_RIS_relay}. Therefore, the large-scale fading between the BS and user $k$ through RISs can be expressed as
\begin{equation}\label{large-scale fading_difuuse}
{L_{{\rm{D}},m,k}} = d_1^{ - {\alpha _1}}d_{m,k}^{ - {\alpha _2}},
\end{equation}
where $\alpha_1$ and $\alpha_2$ denote the path loss exponent of the BS-RIS and RIS-user links, respectively.
We then discuss the minimal required number of RISs for the diffuse scattering scenario in the following Lemma.

\begin{lemma}\label{lemma1:minimum required number of RIS diffuse scattering}
Assuming that the small-scale fading environments of BS-RIS, RIS-user and BS-user links are strong enough, i.e. $\mathcal{K}_1=\mathcal{K}_2 \sim \infty$, where the elements in fading matrixes are all one. Hence, the following constraint in diffuse scattering scenario needs to be met for implementing the proposed SCB design:
\begin{equation}\label{diffuse_fading environment contraint}
{N^2} \ge \frac{{{{(M - 1)}^2}d_{{\rm{b}},m,k}^{ - {\alpha _3}}}}{{d_1^{ - {\alpha _1}}d_{m,k}^{ - {\alpha _2}}}}.
\end{equation}
\begin{proof}
In order to limit the inter-cluster interference, the power level of the reflected signals need to be higher than that of the BS-user links. Thus, the following constraint needs to be met~\cite{Coherence_wave}:
\begin{equation}\label{diffuse_proof_first constraint}
Nd_1^{ - \frac{{{\alpha _1}}}{2}}d_{m,k}^{ - \frac{{{\alpha _2}}}{2}} \ge (M - 1)d_{{\rm{b}},m,k}^{ - \frac{{{\alpha _3}}}{2}}.
\end{equation}
After some algebraic manipulations, the results in~\eqref{diffuse_fading environment contraint} can be readily obtained. Thus, the proof is complete.
\end{proof}
\end{lemma}

2) \emph{Anomalous Reflector Scenario:} We the turn our attention to the anomalous reflector scenario, where the frequency of wave is usually sufficiently high, the large-scale fading can be considered as anomalous reflectors~\cite{renzo_RIS_relay}. Therefore, the large-scale fading between the BS and user $k$ through RISs can be expressed as
\begin{equation}\label{large-scale fading_relector}
{L_{{\rm{A}},m,k}} = {({d_1} + d_{m,k}^{\frac{{{\alpha _2}}}{{{\alpha _1}}}})^{ - {\alpha _1}}}.
\end{equation}
If $\alpha _1=\alpha _2$, the large-scale fading can be simplified to
\begin{equation}\label{large-scale fading_relector_simpli}
{L_{ {\rm{A}},m,k}} = {({d_1} + {d_{m,k}})^{ - {\alpha _1}}}.
\end{equation}

\begin{lemma}\label{lemma2:minimum required number of RIS anomalous reflector}
Let us assume that the small-scale fading environments of BS-RIS, RIS-user and BS-user links are strong enough, i.e. $\mathcal{K}_1=\mathcal{K}_2 \sim \infty$, where the elements in fading matrixes are all one. Hence, the following constraint in anomalous reflector scenario needs to be met for implementing the proposed SCB design:
\begin{equation}\label{anomalous reflector_ environment contraint}
{N^2} \ge \frac{{{{(M - 1)}^2}d_{{\rm{b}},m,k}^{ - {\alpha _3}}}}{{{{({d_1} + d_{m,k}^{\frac{{{\alpha _2}}}{{{\alpha _1}}}})}^{ - {\alpha _1}}}}}.
\end{equation}
\begin{proof}
Similar to {\bf {Lemma~\ref{lemma1:minimum required number of RIS diffuse scattering}}}, and by utilizing~\eqref{large-scale fading_relector}, the results in~\eqref{anomalous reflector_ environment contraint} can be obtained.
\end{proof}
\end{lemma}

Then we conclude the feasibility of two alternatives scenarios in Table~\ref{feasibility_analysis}, and we set $d_1={d_{m,k}}=80$m and $d_{{\rm{b}},m,k}=100$m. $N$ represents the minimal required number of RISs.

\begin{table}
\caption{\\ FEASIBILITY ANALYSIS}
\centering
\begin{tabular}{|l|c|c|c|c|}
\hline
RIS Mode & $\alpha_1$ & $\alpha_2$ & $\alpha_3$ & $N$ \\
\hline
\multirow{3}{*}{Diffuse scattering}
& $3.5$ & $3.5$ & $3.5$ & $1449$\\
\cline{2-5}
& $2.2$ & $3.5$ & $3.5$ & $84$\\
\cline{2-5}
& $2.2$ & $2.2$ & $3.5$ & $5$\\
\hline
\multirow{3}{*}{Anomalous reflector}
& $3.5$ & $3.5$ & $3.5$ & $3$\\
\cline{2-5}
& $2.2$ & $3.5$ & $3.5$ & $1$\\
\cline{2-5}
& $2.2$ & $2.2$ & $3.5$ & $1$\\
\hline
\end{tabular}
\label{feasibility_analysis}
\end{table}

\begin{remark}\label{remark1:application scenario}
The results in \rm{TABLE}~\ref{feasibility_analysis} demonstrate that the proposed SCB design is only applicable for the case of $\alpha_1>\alpha_3$ and $\alpha_2>\alpha_3$ in the diffuse scattering scenario.
\end{remark}

\begin{remark}\label{remark21:Anomalous reflectorscenario}
The results in \rm{TABLE}~\ref{feasibility_analysis} demonstrate that the proposed SCB design can be beneficially implemented in the anomalous reflector scenarios for the case that both the BS-RIS and RIS-user links experience Rayleigh fading channels.
\end{remark}

\subsection{Passive Beamforming Designs}

We first pay our attention to the signal model, and the information bearing vector at the BS can be expressed as:
\begin{equation}\label{information bearing}
{\rm \bf{s}} = \left[ {\begin{array}{*{20}{c}}
{{\alpha _1}{s_{1,1}} +  \cdots  + {\alpha _K}{s_{1,K}}}\\
 \vdots \\
{{\alpha _1}{s_{M,1}} +  \cdots  + {\alpha _K}{s_{M,K}}}
\end{array}} \right],
\end{equation}
where $s_{m,k}$ denotes the signal intended for user $k$ in cluster $m$. $\alpha _k$ represents the power allocation factor for user $k$. Based on the NOMA protocol, we have $\sum\limits_{k = 1}^K {\alpha _k^2 = 1}$.

Without loss of generality, we focus on user $k$ in cluster $m$. Thus, the receiving signal at user $k$ in cluster $m$ is given by
\begin{equation}\label{received user signal}
{y_{m,k}} = \left( {{{\rm \bf{G}}_{m,k}}{\bf{\Phi }}{\rm \bf{H}}\sqrt {{L_{m,k}}}  + {{\rm \bf W}_{m,k}}\sqrt {{L_{b,m,k}}} } \right){\rm \bf P}p{\bf{s}} + {N_0},
\end{equation}
where $p$ denotes the transmit power at the BS, ${\rm \bf P}$ denotes the precoding matrix,\\ ${\rm \bf \Phi}  \buildrel \Delta \over = {\rm{diag}}\left[ {{\beta_1} {\phi _1}, {\beta_2}{\phi _2}, \cdots,{\beta_{N}} {\phi _{N}}} \right]$ denotes both the effective phase shifts and amplitude coefficients by RISs. More specifically, $\beta_n  \in \left( {0,1} \right]$ denotes the amplitude coefficient of RIS element $n$, ${\phi _n} = \exp (j{\theta _n}), j=\sqrt{-1}, \forall n = 1,2 \cdots ,N$. ${\theta _n} \in \left[ {0,2\pi } \right)$ denotes the phase shift by RIS element $n$. Finally, the additive white Gaussian noise (AWGN) is denoted by $N_0$, which is a zero-mean complex circularly symmetric Gaussian variable with variance ${\sigma ^2}$. In order to implement the proposed SCB design, the CSIs are assumed to perfectly known at the RIS controller~\cite{ZhangRui_MISO_beams_discrete_2}.

In practice, user $k$ in cluster $m$ applies a detection vector ${{\rm \bf{v}}_{k,m}}$ to its received signals, therefore the user’s observations is given by:
\begin{equation}\label{after detection vector}
{{\tilde y}_{m,k}} = {\rm \bf{v}}_{k,m}^{{H}}\left( {{{\rm \bf {G}}_{m,k}}{\bf{\Phi }}{\rm \bf {H}}\sqrt {{L_{m,k}}}  + {{\rm \bf W}_{m,k}}\sqrt {{L_{b,m,k}}} } \right){\rm \bf{P}}p{\bf{s}} + {\bf{v}}_{k,m}^{{H}}{N_0}.
\end{equation}
In order to provide a general framework, we assume that the active beamformer weights at the BS obey:
\begin{equation}\label{Precoding matrix design}
{\rm \bf{P}} = {{\rm \bf{I}}_M},
\end{equation}
where ${{\rm \bf{I}}_M}$ represents a $M \times M$ identity matrix. The detection vectors of users can be expressed as all one vector as:
\begin{equation}\label{detection vector design}
{\bf{v}}_{k,m}^{{H}}  = \left[ {\begin{array}{*{20}{c}}
1& \cdots &1
\end{array}} \right].
\end{equation}

We then turn our attention to the passive beamforming design at the RISs, where the phase shifts and reflection amplitude coefficients are jointly manipulated. In this article, the passive beamforming design at RISs mainly focuses on interference cancellation, and hence we first remove the $m$-th column of the matrix ${{\bf{W}}_{m,k}}$ as follows:
\begin{equation}\label{Diffuse_scattering_interference effective matrix}
{{{\bf{\bar W}}}_{m,k}} = \left[ {\begin{array}{*{20}{c}}
{{\rm \bf{w}}_1}& \cdots &{{\rm \bf{w}}_{m - 1}}&{{\rm \bf{w}}_{m + 1}}& \cdots &{{\rm \bf{w}}_M}
\end{array}} \right],
\end{equation}
where ${{\rm \bf {w}}_{m - 1}}$ denotes the $(m-1)$-th column of the matrix ${{\bf{W}}_{m,k}}$. Since the elements in the active beamforming weight and detection vectors are all one, the effective inter-cluster interference of all $MK$ users can be expressed as follows:
\begin{equation}\label{Inter-cluster interference}
{\rm \bf{B}} = \left[ {\begin{array}{*{20}{c}}
{ - {{{\bf{\bar W}}}_{1,1}}{{\rm \bf{1}}_{M - 1}}\sqrt {{L_{{\rm{b}},1,1}}} }\\
 \vdots \\
{ - {{{\bf{\bar W}}}_{1,K}}{{\rm \bf{1}}_{M - 1}}\sqrt {{L_{{\rm{b}},1,K}}} }\\
 \vdots \\
{ - {{{\bf{\bar W}}}_{M,K}}{{\rm \bf{1}}_{M - 1}}\sqrt {{L_{{\rm{b}},M,K}}} }
\end{array}} \right],
\end{equation}
where ${{\rm \bf{1}}_{M - 1}} = {\left[ {1,1, \cdots ,1} \right]^{\rm{T}}}$ denotes an $((M-1)\times 1)$ all one vector.

In order to design the passive beamforming weight at RISs, we first define an effective matrix in the diffuse scattering scenarios by stacking the channel gains of all $M K$ users as follows:
\begin{equation}\label{Effective gain all users}
{{{\rm \bf{\tilde H}}}_{\rm{D}}} = \left[ {\begin{array}{*{20}{c}}
{g_{1,1}^{1,1}{h_{1,1}}\sqrt {{L_{{\rm{D}},1,1}}} }& \cdots &{g_{1,N}^{1,1}{h_{N,1}}\sqrt {{L_{{\rm{D}},1,1}}} }\\
 \vdots & \cdots & \vdots \\
{g_{L,1}^{1,1}{h_{1,M}}\sqrt {{L_{{\rm{D}},1,1}}} }& \cdots &{g_{L,N}^{1,1}{h_{N,M}}\sqrt {{L_{{\rm{D}},1,1}}} }\\
 \vdots & \cdots & \vdots \\
{g_{L,1}^{M,K}{h_{1,M}}\sqrt {{L_{{\rm{D}},M,K}}} }& \cdots &{g_{L,N}^{M,K}{h_{N,M}}\sqrt {{L_{{\rm{D}},M,K}}} }
\end{array}} \right],
\end{equation}
where ${\bf{\tilde H}}_{\rm D}$ is a $(LKM) \times N$ element matrix. Then we define an effective RIS vector $\widetilde {\bf{\Phi }} = {\left[ {{\beta _1}{\phi _1},{\beta _2}{\phi _2}, \cdots ,{\beta _N}{\phi _N}} \right]^{\rm {T}}}$, which is an $N \times 1$ effective vector.
Hence, in order to limit the inter-cluster interference at each user in the diffuse scattering scenarios, the following constraint needs to be met:
\begin{equation}\label{diffuse scattering interference constraint}
{{{\bf{\tilde H}}}_{\rm{D}}}\widetilde {\bf{\Phi }} = {\rm \bf{B}}.
\end{equation}
To achieve the ambitions design objective, the solution of RISs can be given as follows:
\begin{equation}\label{RISs design diffusing}
\widetilde {\bf{\Phi }} = {\bf{\tilde H}}_{\rm{D}}^{ - 1}{\rm \bf {B}}.
\end{equation}

Similarly, let us define an effective matrix for the anomalous reflector scenarios as follows:
\begin{equation}\label{effective channel gain matrix}
{{{\rm \bf{\tilde H}}}_{\rm{A}}} = \left[ {\begin{array}{*{20}{c}}
{g_{1,1}^{1,1}{h_{1,1}}\sqrt {{L_{{\rm{A}},1,1}}} }& \cdots &{g_{1,N}^{1,1}{h_{N,1}}\sqrt {{L_{{\rm{A}},1,1}}} }\\
 \vdots & \cdots & \vdots \\
{g_{L,1}^{1,1}{h_{1,M}}\sqrt {{L_{{\rm{A}},1,1}}} }& \cdots &{g_{L,N}^{1,1}{h_{N,M}}\sqrt {{L_{{\rm{A}},1,1}}} }\\
 \vdots & \cdots & \vdots \\
{g_{L,1}^{M,K}{h_{1,M}}\sqrt {{L_{{\rm{A}},M,K}}} }& \cdots &{g_{L,N}^{M,K}{h_{N,M}}\sqrt {{L_{{\rm{A}},M,K}}} }
\end{array}} \right].
\end{equation}
Then, the RISs can be designed as follows:
\begin{equation}\label{RIS design anomalous}
\widetilde {\bf{\Phi }} = {\bf{\tilde H}}_{\rm{A}}^{ - 1}{\rm \bf{B}}.
\end{equation}

Note that we have $rank({{{\rm \bf{\tilde H}}}_{\rm{A}}}) =rank({{{\rm \bf{\tilde H}}}_{\rm{D}}}) \le MKL $, thus for the case of $N < MKL $, there exists no solution for passive beamforming at RISs, which satisfy the constraints of ${\theta _n} \in \left[ {0,2\pi } \right)$ and ${\beta _n} \in \left( {0,1} \right]$, $\forall n = 1 \cdots N$.

\begin{remark}\label{remark3:global solution minimum requirment}
In order to obtain the solution in~\eqref{RISs design diffusing} and in~\eqref{RIS design anomalous}, the constraint of the number of RISs $N>MKL$ needs to be met, otherwise no solution satisfying $\beta_n  \in \left( {0,1} \right]$ can be obtained.
\end{remark}

\begin{lemma}\label{lemma3: overall minimum required number of RIS both scenarios}
Based on~{\bf Lemma~\ref{lemma1:minimum required number of RIS diffuse scattering}} and~{\bf Lemma~\ref{lemma2:minimum required number of RIS anomalous reflector}} as well as~{\bf Remark~\ref{remark3:global solution minimum requirment}}, the following constraints for both the diffuse scattering and anomalous reflector scenarios need to be met for the proposed SCB design:
\begin{equation}\label{overall N diffuse_fading environment contraint}
{N_{\rm{D}}} \ge \max \left\{ {(M - 1){L_{\rm{D}}},MKL} \right\},
\end{equation}
and
\begin{equation}\label{overall N anomalous reflector contraint}
{N_{\rm{S}}} \ge \max \left\{ {(M - 1){L_{\rm{S}}},MKL} \right\},
\end{equation}
where ${L_{\rm{D}}} = {\left( {\frac{{d_{{\rm{b}},m,k}^{ - {\alpha _3}}}}{{d_1^{ - {\alpha _1}}d_{m,k}^{ - {\alpha _2}}}}} \right)^{\frac{1}{2}}}$ and ${L_{\rm{S}}} = {\left( {\frac{{d_{{\rm{b}},m,k}^{ - {\alpha _3}}}}{{{{({d_1} + d_{m,k}^{\frac{{{\alpha _2}}}{{{\alpha _1}}}})}^{ - {\alpha _1}}}}}} \right)^{\frac{1}{2}}}$.
\end{lemma}

\begin{remark}\label{remark3: minimum requirment in practice}
Since the signals cannot be amplified by RISs, the minimal number of RISs is required to be greater than~\eqref{overall N diffuse_fading environment contraint} and~\eqref{overall N anomalous reflector contraint} for the diffuse scattering and anomalous reflector scenarios, respectively.
\end{remark}

In this article, $K$ users are paired to perform NOMA in each cluster, and thus based on the proposed passive beamforming design at RISs, the signal-to-interference-plus-noise ratio (SINR) of user $k$ in cluster $m$ for the I-RIS cases in both the diffuse scattering and anomalous reflector scenarios can be expressed as
\begin{equation}\label{SINR_ideal_IC}
SIN{R_{k,m}} = \frac{{\left| {{{w}}^{m,k}} \right|^2{L_{b,m,k}}p\alpha _k^2}}{{\sum\limits_{n =  k+1}^{K} {} \left| {{{w}}^{m,k}} \right|^2{L_{b,m,k}}p\alpha _n^2 + L{\sigma ^2}}},
\end{equation}
where $\left| {{{w}}^{m,k}} \right|^2$ denotes the effective channel gain of user $k$ in cluster $m$, which will be evaluated in the next section.

\subsection{RIS Designs for Finite Resolutions}
In most previous research, perfect RIS assumption was assumed, and hence the amplitude coefficients and phase shifts are continuous. However, in practice, the amplitude coefficients and phase shifts rely on the diodes employed at RISs~\cite{ZhangRui_MISO_beams_discrete_2}. Thus the discrete phase shifts were considered~\cite{shuowen_RIS}. Both perfect and imperfect phase shifters were considered in~\cite{RuiZhang_NOMA_OMA,yuanwei_NOMA_RIS}.
In this article, the amplitude coefficients $\beta$ and phase shifts $\theta$ may be not continuous due to the hardware limitations. We then consider an alternative low-cost implementation for applying multi-bit control to RISs, i.e. each diagonal element of $\Phi$ is selected from a set of discrete finite resolutions. It is also worth mentioning that several amplitude dividers may be employed at the RIS elements, whereas the current hardware design of RISs only contains several phase shifters.
As such, the RIS performs a linear mapping based on an equivalent amplitude-coefficient vector as well as phase-shift vector. It is assumed that the phase shifts and amplitude coefficients at RISs take a finite number of discrete values.
The number of bits $b$ is used to indicate the number of phase shift and amplitude coefficient levels $T$ with $T = 2^b$, and hence the discrete phase shift and amplitude coefficient values can be uniformly mapped to the interval $\left[0, 2\pi\right)$ and $\left[ 0,1 \right]$, respectively. Thus, the set of discrete amplitude coefficient as well as phase values at each RIS element can be given by
\begin{equation}\label{Phase_shift_set}
\widehat \theta  = \left\{ {0,\Delta \theta , \cdots (T - 1)\Delta \theta } \right\},
\end{equation}
and
\begin{equation}\label{amplitude coefficient set}
\widehat \beta  = \left\{ {0,\Delta \beta , \cdots (T - 1)\Delta \beta } \right\},
\end{equation}
where $\Delta \theta  = \frac{{2\pi }}{T}$ and $\Delta \beta  = \frac{1}{T}$. The effective phase shifts and amplitude coefficients matrix ${\widehat {\bf{\Phi }}}$ need to be selected from the above two sets in~\eqref{Phase_shift_set} and~\eqref{amplitude coefficient set}. Based on the proposed design at RISs, the interference residue at user $k$ in cluster $m$ can be transformed into
\begin{equation}\label{interference residue}
{\widehat I_{m,k}} = \left\| {{{\bf{G}}_{m,k}}\widehat {\bf{\Phi }}{\bf{H}}\sqrt {{L_{m,k}}} {{\rm \bf{1}}_M} - {{{\bf{\bar W}}}_{m,k}}\sqrt {{L_{b,m,k}}} {{\rm \bf{1}}_{M - 1}}} \right\|_2^2.
\end{equation}
Hence, the SINR of user $k$ in cluster $m$ for the NI-RIS cases can be given by
\begin{equation}\label{non-ideal case SINR}
{\widehat {SINR}_{k,m}} = \frac{{\left| {{{w}}^{m,k}} \right|^2{L_{b,m,k}}p\alpha _k^2}}{{ {\widehat I_{m,k}}p + \sum\limits_{n = k+ 1}^{K} {} \left| {{{w}}^{m,k}} \right|^2{L_{b,m,k}}p\alpha _n^2 + L{\sigma ^2}}}.
\end{equation}

%

\section{Performance Evaluation}
In this section, new channel statistics, OPs, ERs, SE and EE are illustrated in the I-RIS cases.

\subsection{New Channel Statistics}
In this subsection, new channel statistics are derived for the proposed RIS-aided SCB design in a MIMO-NOMA network, which will be used for evaluating the network's performance.
\begin{lemma}\label{lemma3:new state of effective channel gain}
Assuming that the fading channels of the BS-user links follow Rayleigh distribution. The elements of channel gains are independently and identically distributed (i.i.d.). The distribution of the effective channel gain of user $k$ in cluster $m$ for the I-RIS cases can be given by
\begin{equation}\label{New Gamma distribution in Lemma_m-th user in cluster k}
\left| {{{w}}^{m,k}} \right|^2 \sim \Gamma \left( {L,1} \right),
\end{equation}
where $\Gamma \left( \cdot, \cdot \right)$ represents the Gamma distribution.
\begin{proof}
Please refer to Appendix A.
\end{proof}
\end{lemma}
Then the PDF of the effective channel gain can be expressed as
\begin{equation}\label{PDF OP use channel gain}
{f_1}(x) = \frac{{{L^L}{x^{L - 1}}}}{{\Gamma (L)}}{e^{ - Lx}}.
\end{equation}

\subsection{OP and ER}
We first focus on analyzing the OP of both the diffuse scattering and anomalous scattering scenarios. In this article, user $k$ needs to decode the signals of the farer users by SIC technique, i.e. $1$ to $k-1$ users, and hence the OP of user $k$ in cluster $m$ is defined by
\begin{equation}\label{Outage Defination}
\begin{aligned}
{P_{m,k}} = 1 - \prod\limits_{v =  1}^{k - 1} {\mathbb{P}} \left( { {{\log }_2}(1 + SIN{R_{k \to v}}) > {R_v}} \right),
\end{aligned}
\end{equation}
where $R_v$ denotes the target rate of user $v$ in cluster $m$, and $SIN{R_{k \to v}} = \frac{{{{\left| {{w^{m,k}}} \right|}^2}{L_{b,m,k}}p\alpha _v^2}}{{\sum\limits_{q = v + 1}^K {} {{\left| {{w^{m,k}}} \right|}^2}{L_{b,m,k}}p\alpha _q^2 + L{\sigma ^2}}}$.

Then the OP of user $k$ in cluster $m$ is given in the following Theorem.
\begin{theorem}\label{Theorem1:Outage W user closed form by incomlete gamma}
\emph{Let us assume that $ \alpha _v^2 - \left( {\sum\limits_{q = v + 1}^K {\alpha _q^2} } \right){\varepsilon _v} > 0$ with $v = 1, \cdots ,k$, and the number of RISs obeys the constraints in~{\bf {Lemma~\ref{lemma1:minimum required number of RIS diffuse scattering}}} as well as {\bf{Lemma~\ref{lemma2:minimum required number of RIS anomalous reflector}}}, the closed-form OP expression of user $k$ in cluster $m$ for the I-RIS cases can be expressed as}
\begin{equation}\label{outage analytical results W in theorem1}
\begin{aligned}
{P_{m,k}} = \frac{{\gamma \left( {L,{I_{m,k*}}} \right)}}{{\Gamma (L)}},
\end{aligned}
\end{equation}
\emph{where ${I_{m,k*}} = \max \left\{ {{I_{m,1}}, \cdots {I_{m,k}}} \right\}$, ${I_{m,v}} = \frac{L{{\varepsilon _v}{\sigma ^2}{L_{{\rm{b}},m,k}}}}{{p\left( {\alpha _v^2 - \left( {\sum\limits_{q = v + 1}^K {\alpha _q^2} } \right){\varepsilon _v}} \right)}}$, ${\varepsilon _v} = {2^{{R_v}}} - 1$, and $\gamma (,)$ represents the lower incomplete Gamma function.}
\begin{proof}
Recall that user $k$ needs to decode the signal from user $1$ to user $k-1$ one by one, and based on the OP defined in~\eqref{Outage Defination}, the coverage probability can be written as
\begin{equation}\label{Coverage_probability}
{P_{m,k,{\rm{C}}}} =1-\mathbb{P} \left( {\left| {{{w}}^{m,k}} \right|^2 > {I_{m,1}}, \cdots ,\left| {{{w}}^{m,k}} \right|^2 > {I_{m,k}}} \right).
\end{equation}
By applying ${I_{m,k*}} = \max \left\{ {{I_{m,1}}, \cdots {I_{m,k}}} \right\}$, the coverage probability can be further transformed into
\begin{equation}\label{coverage probability expression}
{P_{m,k,{\rm{C}}}} =1 - \frac{{\gamma \left( {L,{I_{m,k*}}} \right)}}{{\Gamma (L)}}.
\end{equation}
Hence, the results in~\eqref{outage analytical results W in theorem1} can be obtained.
\end{proof}
\end{theorem}

We then focus on the diversity orders of user $k$ in cluster $m$, which can be obtained for evaluating the slope of OP.
\begin{proposition}\label{proposition1: user k in cluster m diversity order}
\emph{From \textbf{Theorem~\ref{Theorem1:Outage W user closed form by incomlete gamma}}, the diversity orders for the I-RIS cases can be determined by expanding the lower incomplete Gamma function, and the diversity order of user $k$ in cluster $m$ of the proposed RIS-aided SCB design can be given by}
\begin{equation}\label{diversity order of user k in cluster m}
{d_{m,k}} =  - \mathop {\lim }\limits_{\frac{{{p}}}{{{\sigma ^2}}} \to \infty } \frac{{\log {P_{m,k}}}}{{\log \frac{{{p}}}{{{\sigma ^2}}}}} \approx L,
\end{equation}
\begin{proof}
Please refer to Appendix B.
\end{proof}
\end{proposition}

\begin{remark}\label{remark4:impact of RAs}
Based on results in~\eqref{diversity order of user k in cluster m}, it is indicated that the diversity orders of all the NOMA users can be approximated to the number of RAs $L$ for the I-RIS cases when the number of RISs is high enough.
\end{remark}

We then turn our attention to the ER of user $K$ in cluster $m$, which is a salient metric for performance analysis, and hence the approximated ER expressions for user $K$ in cluster $m$ is given in the following Theorem.
\begin{theorem}\label{theorem3:ergodic rate W-th user}
\emph{When the number of RISs $N$ is sufficiently high, and $ \alpha _v^2 - \left( {\sum\limits_{q = v + 1}^K {\alpha _q^2} } \right){\varepsilon _v} > 0$ with $v = 1, \cdots ,k$, the ER of user $K$ in cluster $m$ can be expressed in the closed-form as follows:}
\begin{equation}\label{asympto W-th erogodic rate in corollary4}
\begin{aligned}
{R_{m,K}} = \frac{1}{{\ln \left( 2 \right)}}\sum\limits_{i = 0}^{L - 1} {\frac{{{C^i}}}{{i!}}} \left( {\exp (C)Ei( - C) + \sum\limits_{a = 1}^i {{{( - 1)}^{a - 1}}(a - 1)!{{C}^a}} } \right),
\end{aligned}
\end{equation}
where $C = \frac{L{{\sigma ^2}}}{{{p}\alpha _K^2}}$.
\begin{proof}
Please refer to Appendix C.
\end{proof}
\end{theorem}

Furthermore, the SINR of user $k$ in cluster $m$ may approach $SIN{R_{m,k}} = \frac{{\alpha _k^2}}{{\sum\limits_{q = k + 1}^K {\alpha _q^2} }}$ in the high-SNR regimes~\cite{Hou_Single_UAV} based on the SINR analysis in~\eqref{SINR_ideal_IC}. Hence, the expected rate of user $k$ for $k=1 ,\cdots ,K-1$ can be written as ${R_{m,k}} = {\log _2}\left( {1 + \frac{{\alpha _k^2}}{{\sum\limits_{q = k + 1}^K {\alpha _q^2} }}} \right)$, which is a constant.

The high-SNR slope is defined as the asymptotic slope of the logarithmic plot of ER against the transmit power in dBm, which is a key parameter determining the ER in the high-SNR regimes, and hence the high-SNR slope can be expressed as
\begin{equation}\label{High SNR slope of m}
\Delta_{m,k}  = -\mathop {\lim }\limits_{ \Xi  \to \infty } \frac{{{R_{m,k}}}}{{{{\log }_2}\left( {1 + \Xi  } \right)}},
\end{equation}
where $\Xi =\frac{p}{\sigma^2}$.

\begin{proposition}\label{proposition4: m high SNR slopes in massive LIS}
\emph{By substituting~\eqref{asympto W-th erogodic rate in corollary4} into~\eqref{High SNR slope of m}, the high-SNR slope of the nearest user can be given by}
\begin{equation}\label{high SNR slope in massive LIS}
\Delta_{m,k} = 1.
\end{equation}
\end{proposition}

\begin{remark}\label{remark6:impact of ergodic rate in massive LIS}
Based on the results in~\eqref{high SNR slope in massive LIS}, one can know that the high-SNR slopes of user $K$ in each cluster of the I-RIS cases are one, which is not a function of the number of RISs.
\end{remark}

\begin{remark}\label{remark5:high SNR slope other users}
Based on the SINR analysis in~\eqref{non-ideal case SINR} and insights from~\cite{Hou_Single_UAV}, the high-SNR slopes of the $1, \cdots ,K-1$-th NOMA users are 0 in both the I-RIS and NI-RIS cases.
\end{remark}

\begin{remark}\label{remark5:impact of non-ideal-diversity order}
Based on the SINR analysis in~\eqref{non-ideal case SINR}, the diversity orders and high-SNR slopes of all the NOMA users are 0 in the NI-RIS cases.
\end{remark}

We then compare the OP of the proposed SCB enhanced MIMO-NOMA network and its OMA counterparts in the following Corollary, i.e. TDMA. The OMA counterparts adopted in this article is that by dividing the $K$ users in equal time slots.

\begin{corollary}\label{theorem of OMA OP_benchmark}
Let us assume that multiple users are divided in equal time slots in the OMA counterparts, the closed-form OP expression of user $k$ in cluster $m$ of the I-RIS cases is given by:
\begin{equation}\label{theorem of OMA terrestrial fomula}
\begin{aligned}
{{\bar P}_{m,k}} = \frac{{\gamma \left( {L,{I_{m,k,O}}} \right)}}{{\Gamma (L)}},,
\end{aligned}
\end{equation}
\emph{where ${I_{m,k,O}} =  \frac{L{{\varepsilon _O}{\sigma ^2}{L_{{\rm{b}},m,k}}}}{{p}}$, and ${\varepsilon _O} = {2^{{K{R_v}}}} - 1$.}
\begin{proof}
We first express the OP of user $k$ in cluster $m$ in the OMA counterparts as follows
\begin{equation}\label{OMA SINR expression}
\mathbb{P} \left\{ {\frac{1}{K}lo{g_2}\left( {1{\rm{ + }}SN{R_{m,k,O}}} \right) > {R_{m,k}}} \right\},
\end{equation}
where $SN{R_{m,k,O}} = \frac{{p{L_{{\rm{b}},m,k}}{{\left| {{w^{m,k}}} \right|}^2}}}{{L{\sigma ^2}}}$. Similar to Theorem~\ref{Theorem1:Outage W user closed form by incomlete gamma}, the results in~\eqref{theorem of OMA terrestrial fomula} can be obtained.
\end{proof}
\end{corollary}

\subsection{SE and EE}

Here, we focus on the SE of cluster $m$, which can be formulated based on the ER analysis in the previous subsection.
\begin{proposition}\label{proposition5: spectrum efficiency}
\emph{In the proposed SCB design, the SE of cluster $m$ can be given by }
\begin{equation}\label{spectrum efficiency}
S_m=\sum\limits_{k = 1}^K {{R_{m,k}}} .
\end{equation}
\end{proposition}

Since RISs are passive equipment, where only RIS controller needs power supply~\cite{EE_model_massive_MIMO,glob_energy_model,energy_model_LIS}, hence we model the total dissipation power of the proposed SCB design as
\begin{equation}\label{total energy massive LIS}
{P_{e}} = {P_{{\rm{B,s}}}} + K P_{\rm{U}} + p{{{\varepsilon _b}}} + {N}{P_L},
\end{equation}
where ${P_{{\rm{B,s}}}}$ and ${{\varepsilon _b}}$ denote the power consumption and the efficiency of power amplifier at the BS, respectively. $P_{\rm{U}}$ and ${P_L}$ denote the power consumption of each user and each RIS controller, respectively. Hence, the EE of the proposed design is given by the following Proposition.

\begin{proposition}\label{proposition7: energy efficiency massive LIS}
\emph{The EE of cluster $m$ in the proposed SCB design can be given by}
\begin{equation}\label{energy efficiency massive LIS}
{\Theta _{EE}} = \frac{S_m}{{{P_{e}}}},
\end{equation}
where $S_m$ and $P_{e}$ are obtained from~\eqref{spectrum efficiency} and~\eqref{total energy massive LIS}, respectively.
\end{proposition}

\section{Numerical Results}

In this section, numerical results are provided for the performance evaluation of the proposed SCB design.
Monte Carlo simulations are provided for verifying the accuracy of our analytical results. The transmission bandwidth of the proposed network is set to $BW=100$ MHz. In practice, the power of the AWGN is related to the bandwidth, which can be modeled as $\sigma^2= −-174+ 10{\rm log}_{10}(BW)$ dBm. For simplicity, the number of TAs is set to $M=2$, and the number of users is set to $K=2$. Based on NOMA protocol, the paired users share the power with the power allocation factors $\alpha_1^2=0.6$ and $\alpha_2^2=0.4$.
The fading factors are set to $\mathcal{K}_1=\mathcal{K}_{2,1}=\mathcal{K}_{2,2}=3$. The distance of the BS-RIS links is set to $d_1=80$m, and the distance of the RIS-user links are set to $d_{1,2}=80$m and $d_{1,1}=160$m and those of the BS-user links are set to $d_{b,1,2}=100$m and $d_{b,1,1}=200$m. The path loss exponents of the BS-user links are set to $\alpha_3=3.5$ while those of the BS-RIS and RIS-user links are set to $\alpha_1=\alpha_2=2.2$. The target rates are $R_2=1.5$ and $R_1=1$ bits per channel use (BPCU), unless otherwise clarified.

\begin{figure}[t!]
\centering
\includegraphics[width =5in]{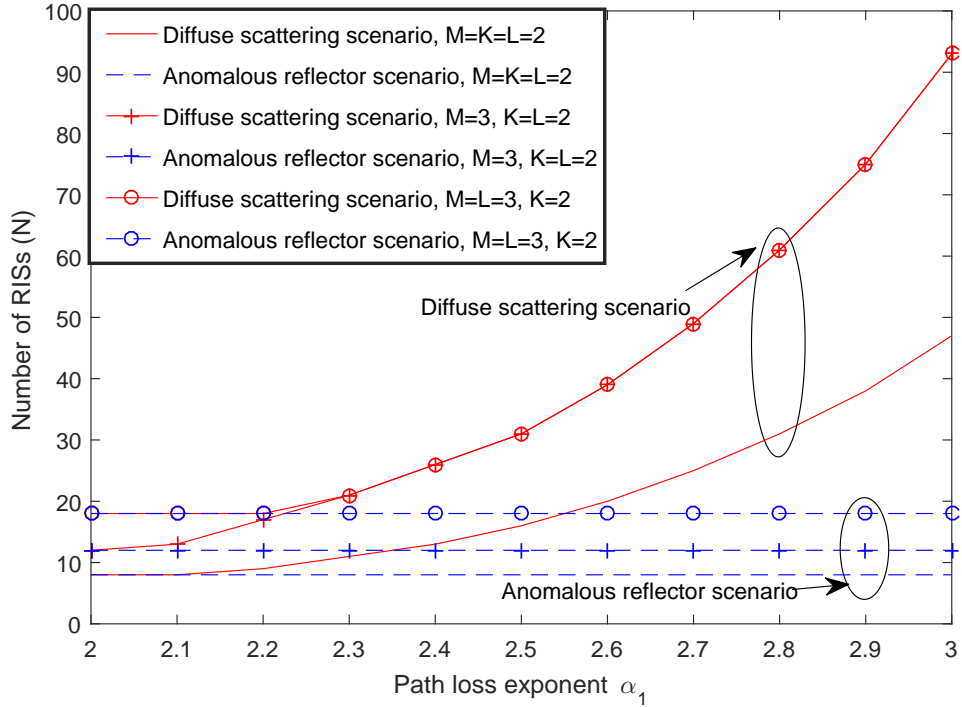}
\caption{Minimal required number of RISs for the I-RIS cases versus the path loss exponent.}
\label{Minimum_number of RIS_fig2}
\end{figure}

\emph{1) Minimal Required Number of RISs:} In Fig.~\ref{Minimum_number of RIS_fig2}, we evaluate the minimal required number of RISs for implementing the proposed SCB design. On the one hand, observe that in the anomalous reflector scenario, the minimal required number of RISs is only impacted by the number of TAs, RAs as well as the number of users. On the other hand, we can see that as the path loss exponent increases, the minimal required number of RISs increases for the diffuse scattering scenarios. This phenomenon indicates that the LoS links of both the BS-RIS and RIS-user links are required for implementing the proposed SCB design in the diffuse scattering scenarios, whereas the LoS links are not necessary for the anomalous reflector scenarios. Observe that for the case of $M=3, K=L=2$ as well as $M=L=3, K=2$, the minimal required RISs are identity, which indicates that the diffuse scattering scenario is more susceptible to the path loss exponent of both the BS-RIS and RIS-user links.

\begin{figure}[t!]
\centering
\includegraphics[width =5in]{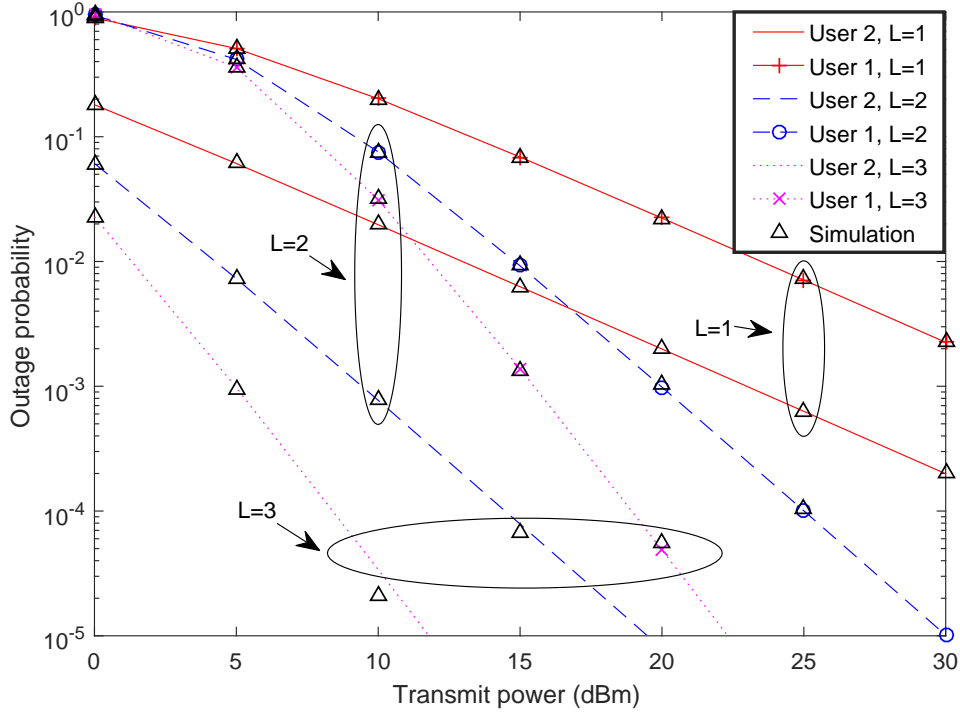}
\caption{OP of the I-RIS cases versus the transmit power with different number of RAs. The analytical results are calculated from~\eqref{outage analytical results W in theorem1}. The number of RISs is set to $N=5 N_D$.}
\label{ideal-RIS case_fig3}
\end{figure}

\emph{2) Impact of the Number of RAs:} In Fig.~\ref{ideal-RIS case_fig3}, we evaluate the OP of the proposed SCB design in the RIS-aided MIMO-NOMA networks. We can see that as the number of RAs equipped at each user increases, the OP decreases. There are two reasons, where 1) since the precoding matrix is an identity matrix, and the detection vector is an all one vector, the RISs are capable of beneficially eliminating the inter-cluster interference in the I-RIS cases; 2) the received signal power can be significantly increased as more RAs are employed.
One can observe that the slopes of the curves are approximated to the number of RAs, which validates our~\textbf{Remark~\ref{remark4:impact of RAs}}.

\begin{figure}[t!]
\centering
\includegraphics[width =5in]{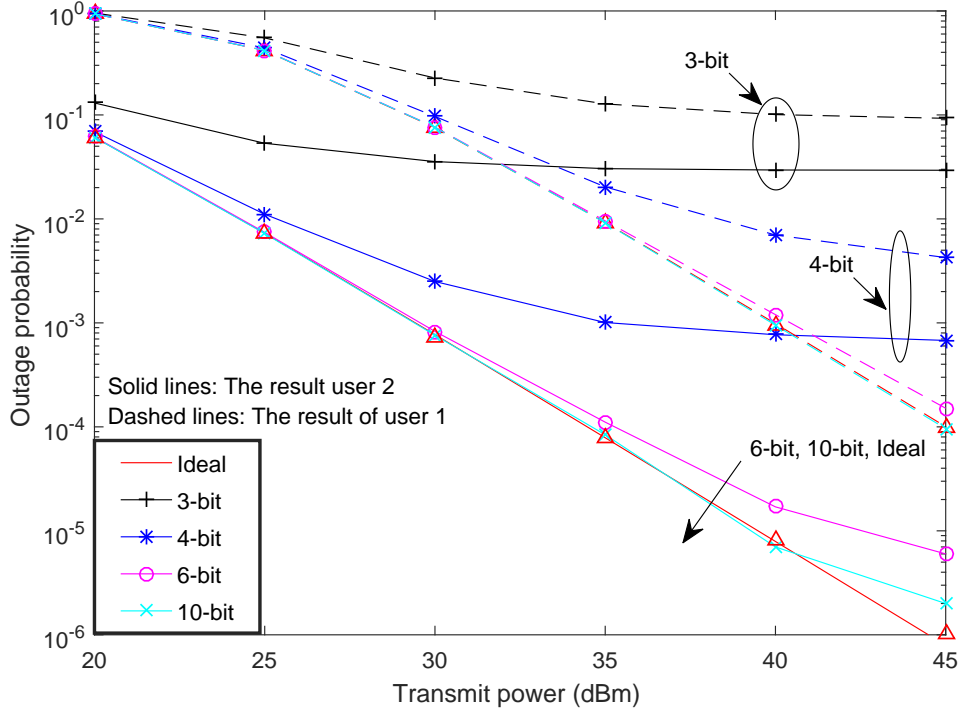}
\caption{OP of the NI-RIS cases versus the transmit power with different number of bits. The number of RISs is set to $N=5 N_D$.}
\label{non-ideal-RIS case fig4}
\end{figure}

\emph{3) Impact of the Number of Bits:} In Fig.~\ref{non-ideal-RIS case fig4}, we evaluate the OP of the paired NOMA users in the different number of bits, where the OP of the paired NOMA users in the I-RIS cases is provided as the benchmark schemes. We can see that as the transmit power increases, the OP floors occur. Observe that as the number of bits increases from 3-bit to 6-bit, the OP can be beneficially decreased. This is due to the fact that the higher number of bits is capable of increasing the resolution of each RIS element. It is also worth noting that 6-bit resolution is enough for obtaining the near-minimal OP, which indicates that the minimized OP is obtainable by appropriate setting the number of bits. Based on the simulation results, the diversity orders of the NI-RIS cases are zero, which verifies the insights gleaned from~\textbf{Remark~\ref{remark5:impact of non-ideal-diversity order}}.

\begin{figure}[t!]
\centering
\includegraphics[width =5in]{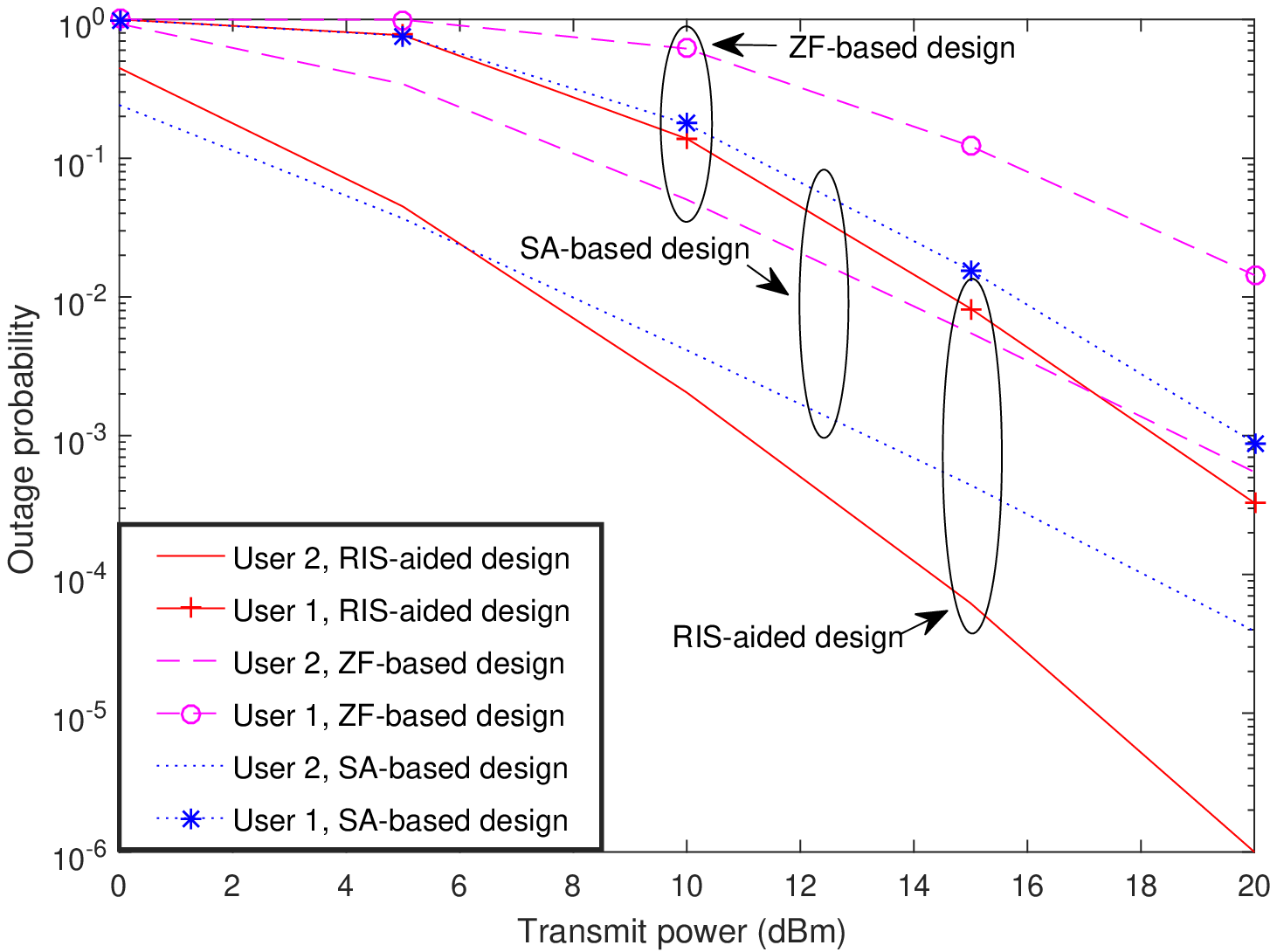}
\caption{OP of both the RIS-aided, ZF-based as well as SA-based designs versus the transmit power. The number of RAs is set to $L=3$. The lengths of the RIS-user links are set to $d_{1,2}=130$m and $d_{1,1}=260$m and those of the BS-user links are set to $d_{b,1,2}=150$m and $d_{b,1,1}=240$m. The number of RISs is set to $N=5 N_D$.}
\label{compare_OP_ZF and SA}
\end{figure}

\emph{4) Comparing ZF-based and SA-based Designs:} In Fig.~\ref{compare_OP_ZF and SA}, combined with the insights inferred from~\cite{zero-forcing_detection,signal-alignment-design}, we compare the OP of the paired NOMA users in the proposed RIS-aided SCB design, ZF-based design as well as SA-based design. In order to provide further engineering insights, we consider that the phase shifts and amplitude coefficients of RISs can be perfectly manipulated. On the one hand, one can observe that the proposed RIS-aided design is capable of outperforming both the classic ZF-based and SA-based designs. On the other hand, the diversity order of the proposed RIS-aided design is $L$, which is higher than both the classic ZF-based and SA-based designs, and hence illustrate the benefits of the proposed RIS-aided SCB design. The detail of diversity orders is concluded in TABLE~\ref{RIS and ZfSA}.

\begin{table}
\caption{\\ COMPARISON BETWEEN RIS-AIDED SCB, ZF-BASED AND SA-BASED DESIGNS. ``MRN'' DENOTES ``MINIMAL REQUIRED NUMBER''.}
\centering
\begin{tabular}{|l|c|c|c|}
\hline
Mode & TAs & MRN of RAs & Antenna Gain \\
\hline
ZF-based~\cite{zero-forcing_detection} & $M$ & $L \ge M$ & $L-M+1$\\
\hline
SA-based~\cite{signal-alignment-design} & $M$ & $L \ge \frac{M}{2}+1$ & $L$\\
\hline
RIS-aided SCB design & $M$ & $L \ge 1$ & $L$\\
\hline
\end{tabular}
\label{RIS and ZfSA}
\end{table}

\begin{figure}[t!]
\centering
\includegraphics[width =5in]{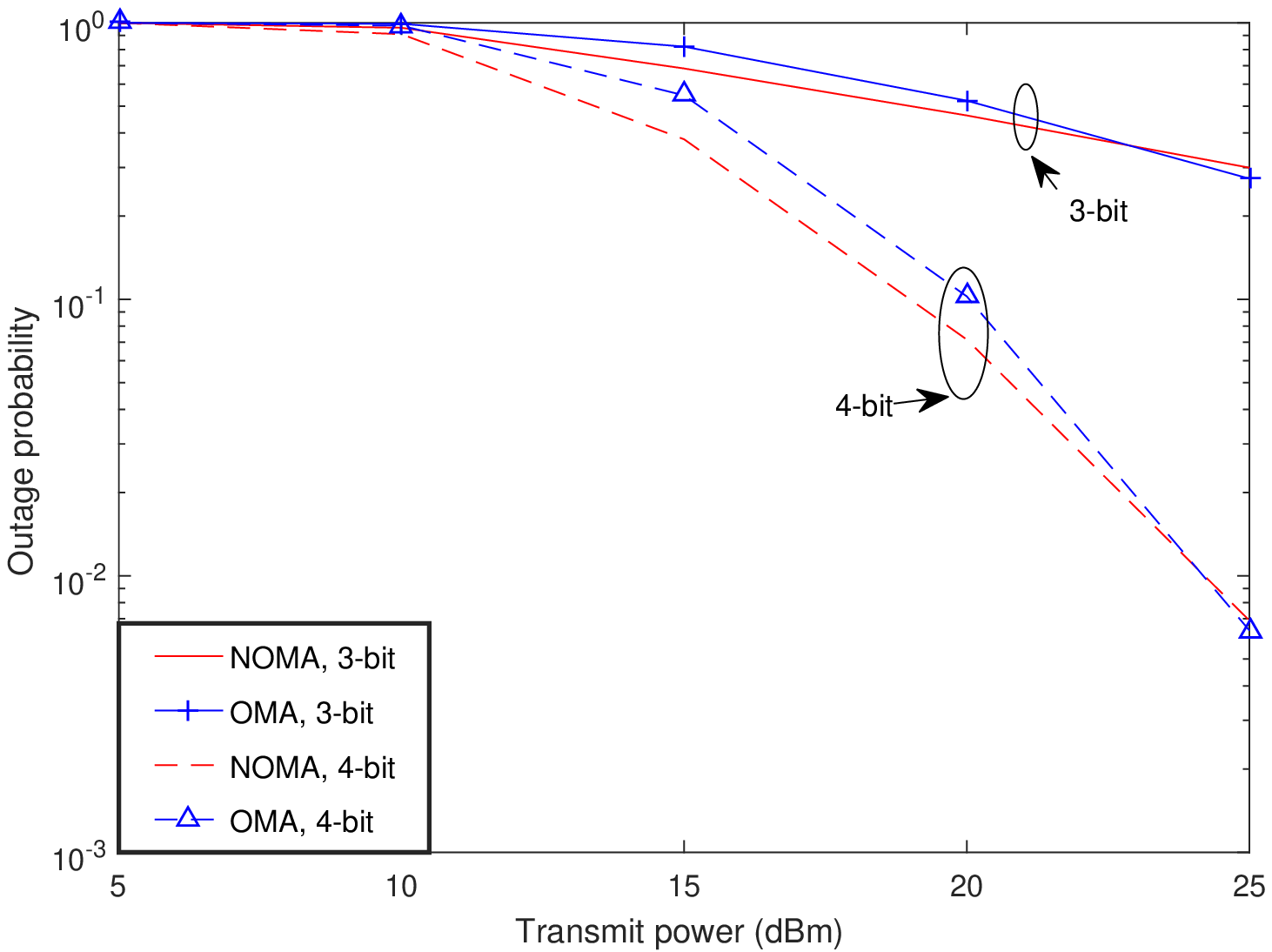}
\caption{OP of the NI-RIS cases in both the NOMA and OMA networks with different number of bits. The number of RISs is set to $N=8 N_D$.}
\label{non-ideal-RIS case OMA_NOMA}
\end{figure}

\emph{5) Comparing RIS-aided NOMA and OMA networks:} In Fig.~\ref{non-ideal-RIS case OMA_NOMA}, we evaluate the OP of both the RIS-aided NOMA and OMA networks. The OPs of NOMA and OMA networks are derived by $P_{\rm NOMA}=P_{m,1} \times P_{m,2}$ and $P_{\rm OMA}={\bar P}_{m,1} \times {\bar P}_{m,2}$, respectively. As can be seen from Fig.~\ref{non-ideal-RIS case OMA_NOMA}, the OP of the RIS-aided NOMA networks is lower than that of the RIS-aided OMA networks, which implies that RIS-aided NOMA network is capable of providing better network performance than its OMA counterpart. Observed that for both the 3-bit and 4-bit resolutions, an optimal point exists due to the fact that there is a cross point of curves in the proposed SCB design. This indicates that the RIS-aided hybrid NOMA/OMA networks may be a good solution.

%
%

\begin{figure}[t!]
\centering
\includegraphics[width =5in]{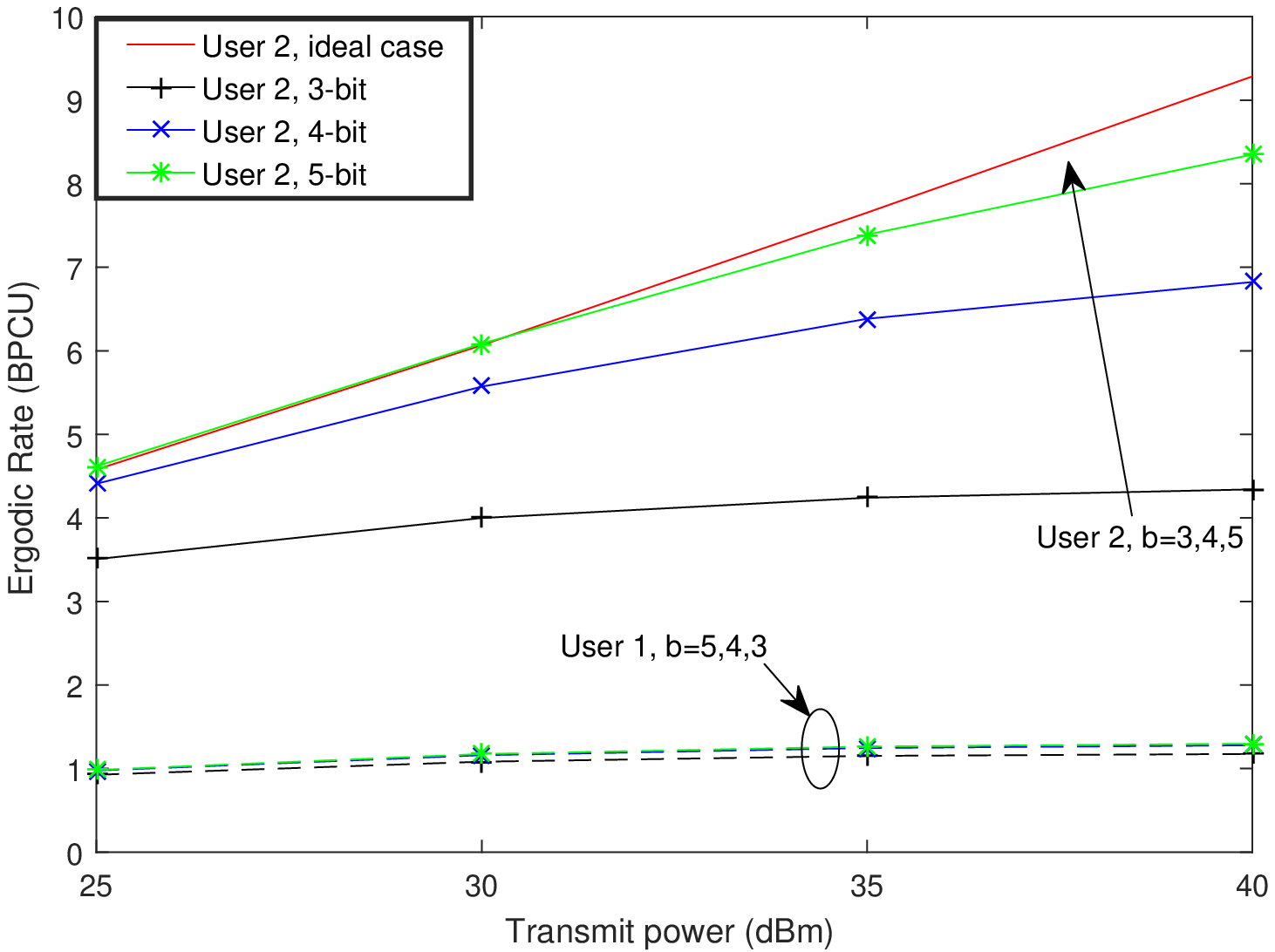}
\caption{ER of the NI-RIS cases versus the number of RISs with different number of bits. The number of RISs is set to $N=5 N_D$.}
\label{non-ideal-RIS case ER impact on N bits}
\end{figure}

\emph{6) Impact of the Number of Bits on ER:} We then evaluate the ER of the paired NOMA users versus the transmit power with the different number of bits in Fig.~\ref{non-ideal-RIS case ER impact on N bits}. Observe that as the transmit power increases, the ER ceilings occur in the NI-RIS cases. One can also observe that as the number of bits increases, the ER gaps between the I-RIS and NI-RIS cases are getting smaller. One can also observe that for the case of $b=5$, the ER of both the NI-RIS and I-RIS cases are nearly identical, which indicates that the 5-bit finite resolution is high enough for the proposed RIS-aided SCB design. It is also worth noting that the ER of user 1 in cluster $m$ is constant with the different number of bits, which also verifies the insights gleaned from~\textbf{Remark~\ref{remark5:high SNR slope other users}}. Observed that for the 3-bit and 4-bit resolutions, there exists an optimal number of RISs for maximizing the SE. This phenomenon also indicates that the RISs can be activated appropriately for enhancing the network's SE in the NI-RIS cases.

\begin{figure}[t!]
\centering
\includegraphics[width =5in]{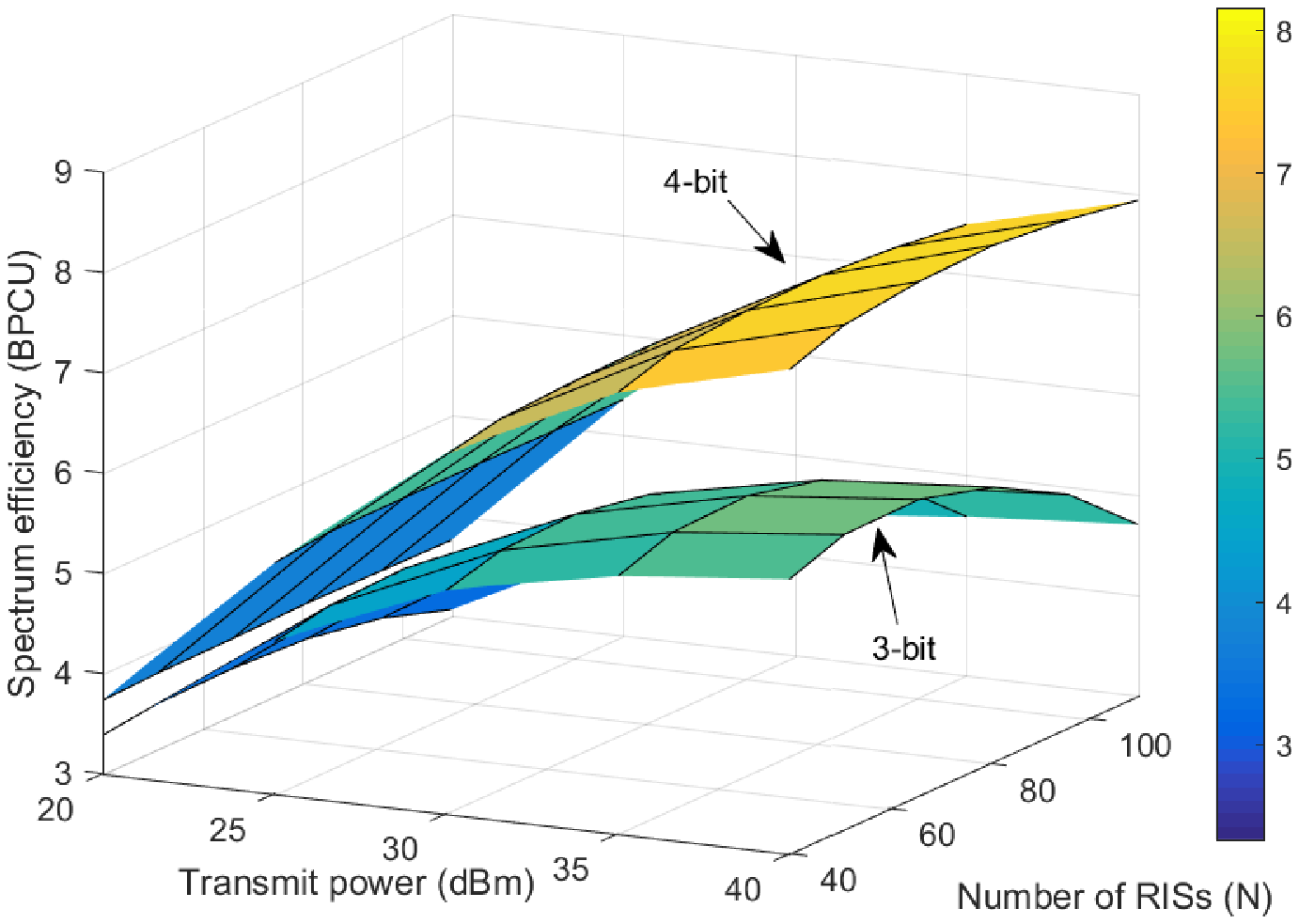}
\caption{SE of the NI-RIS cases versus the number of RISs with different number of bits.}
\label{non-ideal-RIS case SE 3D}
\end{figure}

\emph{7) Spectrum Efficiency:} In Fig.~\ref{non-ideal-RIS case SE 3D}, we evaluate the SE of cluster $m$ of the proposed RIS-aided SCB design. On the one hand, observe that the SE improves as the transmit power increases. However, as the increase of the number of RISs, observe that the slope of the SE is negative of the 3-bit cases, which indicates that there exists an optimal value of the number of RISs that maximizes the SE. It is also worth noting that the SE of the 4-bit cases are nearly identical compared to the 3-bit cases, which indicates that the RIS-aided SCB design with 4-bit finite resolutions becomes more competitive.

\begin{figure}[t!]
\centering
\includegraphics[width =5in]{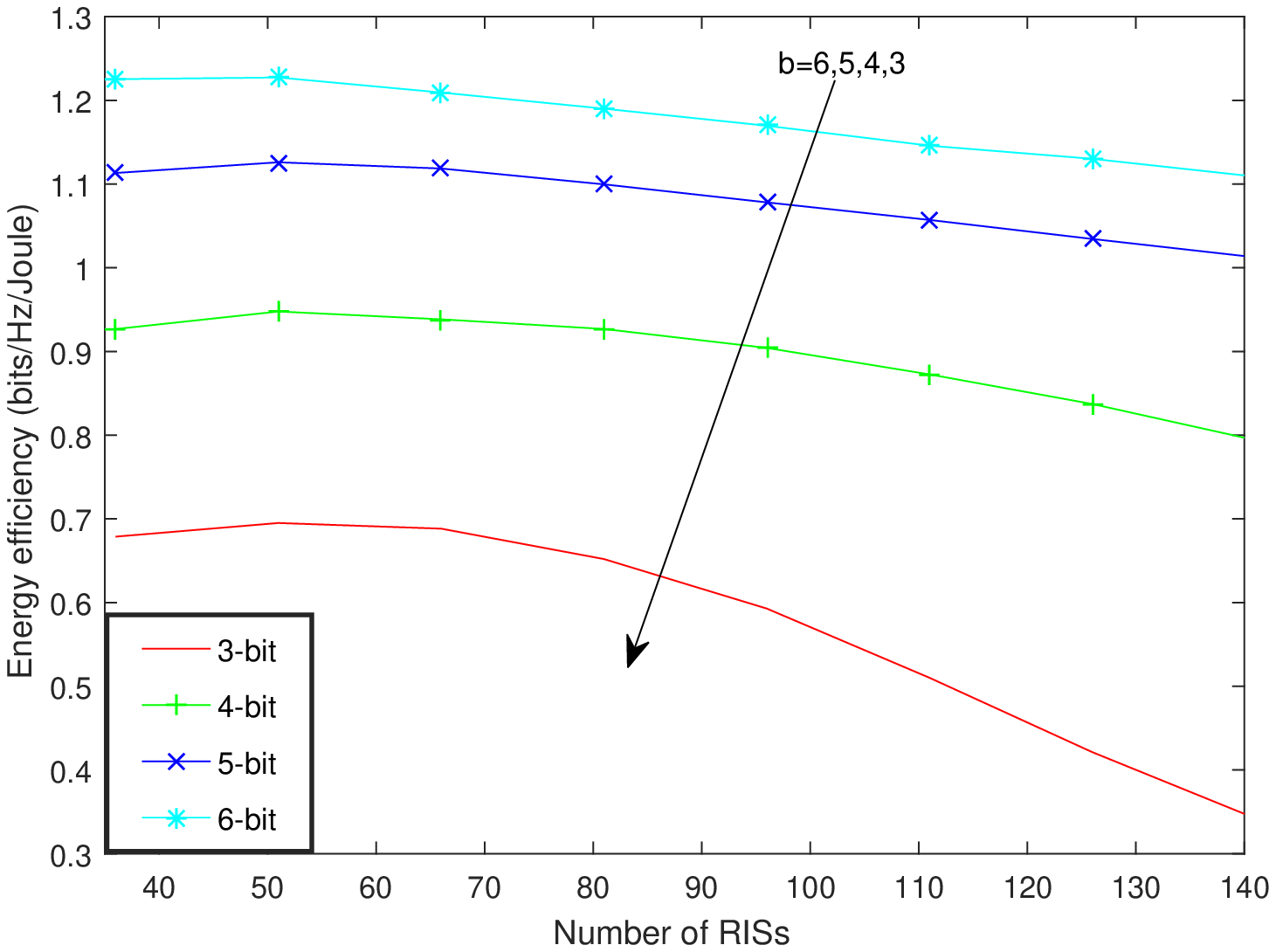}
\caption{EE of the NI-RIS cases versus the number of RISs with different number of bits.}
\label{non-ideal-RIS case EE}
\end{figure}

\emph{8) Energy Efficiency:} We then evaluate the EE of the proposed SCB design in Fig.~\ref{non-ideal-RIS case EE}. Several observations can be concluded as: 1) One can observe that the slopes of EE are negative in the NI-RIS cases. This is due to the fact that in the proposed SCB design, the main goal of passive beamforming is to eliminate the inter-cluster interference, and hence the maximum EE is fixed, which leads to the low EE when the number of RISs is too high; 2) The EE increases as the number of bits increase, this is due to the fact that the inter-cluster interference residue decreases when there are more potential solutions at RISs; and 3) Based on the fact that the EE curve decreases, it is readily to observe that there exists an optimal value of the number of RISs.


\section{Conclusions}

We first reviewed previous contributions related to the RIS-aided SEB designs in this article, and we then proposed a novel RIS-aided SCB design. In order to provide a general design of RIS networks, we adopted a MIMO-NOMA network, where the passive beamforming weights of the RISs were designed. The channel statistics, OPs, ERs, SE as well as EE were derived in closed-form for characterizing the system performance. Compared to the previous ZF-based and SA-based designs, the proposed RIS-aided SCB design releases the constraint of the number of RAs.
An important future direction is to minimize the required number of RISs and RAs by jointly designing both the active beamforming, passive beamforming, and detection vectors.

\numberwithin{equation}{section}
\section*{Appendix~A: Proof of Lemma~\ref{lemma3:new state of effective channel gain}} \label{Appendix:As}
\renewcommand{\theequation}{A.\arabic{equation}}
\setcounter{equation}{0}

Recall that the effective channel vector ${{\rm \bf{w}}_m^{m,k}}$ contains $L$ elements, and by utilizing the proposed detection vector at the users, we can obtain the following channel gain:
\begin{equation}\label{effective channel gain first}
\left | {{{w}}^{m,k}} \right|^2 = \sum\limits_{l = 1}^L {{{\left| {w_{l,1}^{m,k}} \right|}^2}} .
\end{equation}

Based on the result derived in~\eqref{effective channel gain first}, and exploiting the fact that the elements of $\left| {{\rm \bf{W}}_{m,k}} \right|$ are i.i.d., the mean and variance of the effective channel gain can be given by using the property of random variables as follows
\begin{equation}\label{appendix first mean}
{E_1} = \sum\limits_{l = 1}^L {\mathbb{E}} \left( {{{\left| {w_{l,1}^{m,k}} \right|}^2}} \right) = L,
\end{equation}
and
\begin{equation}\label{appendix first variance}
\begin{aligned}
{V_1} = \sum\limits_{l = 1}^L {\mathbb{V}} \left( {{{\left| {w_{l,1}^{m,k}} \right|}^2}} \right) = L.
\end{aligned}
\end{equation}
Thus, the effective channel gain can be rewritten as
\begin{equation}\label{appendix channel multiple distribution}
\left| {{{w}}^{m,k}} \right|^2 \sim \left( {L,L} \right).
\end{equation}
Hence, the effective channel gain can be obtained in a more elegant form in~\eqref{New Gamma distribution in Lemma_m-th user in cluster k}.

\numberwithin{equation}{section}
\section*{Appendix~B: Proof of Proposition~\ref{proposition1: user k in cluster m diversity order}} \label{Appendix:Bs}
\renewcommand{\theequation}{B.\arabic{equation}}
\setcounter{equation}{0}

In order to glean the diversity order of the proposed SCB design, the lower incomplete Gamma function can be expanded as follows~\cite{Table_of_integrals}:
\begin{equation}\label{Appendix B inGamma first expansion}
\gamma \left( {L,{I_{m,k*}}} \right) = {\rm{ }}\sum\limits_{s = 0}^\infty  {} \frac{{\Gamma \left( L \right)}}{{\Gamma \left( {L + s + 1} \right)}}{\left( {{I_{m,k*}}} \right)^{L + s}}\exp \left( { - {I_{m,k*}}} \right).
\end{equation}
When replacing $\exp \left( { - {I_{m,k*}}} \right)$ by its power series expansion, the OP can be further transformed into
\begin{equation}\label{Appendix B OP first}
{{\bar P}_{m,k}} = {\left( {{I_{m,k*}}} \right)^L}\sum\limits_{s = 0}^\infty  {} \frac{{{{\left( { - {I_{m,k*}}} \right)}^s}}}{{s!\left( {L + s} \right)}}.
\end{equation}

Thus, by applying the definition of diversity order, the results in~\eqref{diversity order of user k in cluster m} can be gleaned, and the proof is complete.

\numberwithin{equation}{section}
\section*{Appendix~C: Proof of Theorem~\ref{theorem3:ergodic rate W-th user}} \label{Appendix:Cs}
\renewcommand{\theequation}{B.\arabic{equation}}
\setcounter{equation}{0}

The proof starts by expressing the ER of user $K$ in cluster $m$ as follows:
\begin{equation}\label{Appendix D first define}
\begin{aligned}
{R_{m,K}} &= \mathbb{E} \left\{ {{{\log }_2}\left( {1 + SIN{R_{m,K}}\left( x \right)} \right)} \right\} \\
& =  - \int\limits_0^\infty  {{{\log }_2}(1 + x)} d\left( {1 - F\left( x \right)} \right)\\
& = \frac{1}{{\ln \left( 2 \right)}}\int\limits_0^\infty  {\frac{{1 - F\left( x \right)}}{{1 + x}}} dx.
\end{aligned}
\end{equation}

The cumulative distribution function of user $K$ in cluster $m$ can be calculated as
\begin{equation}\label{Appendix D CDF define}
F\left( x \right) = \left( {\frac{{\gamma \left( {L,Cx} \right)}}{{\Gamma ({L})}}} \right).
\end{equation}

Based on expression in~\eqref{Appendix D first define}, the lower incomplete Gamma function ought to be expanded with a constant for obtaining the closed-form expressions, hence we expand the lower incomplete Gamma function as follows:
\begin{equation}\label{Appendix D lower incomplete gamma function expansion}
\frac{{\gamma \left( {L,Cx} \right)}}{{\Gamma (L)}}{\rm{ = }}1{\rm{ - }}\sum\limits_{i = 0}^{L - 1} {\frac{{{{\left( {Cx} \right)}^i}}}{{i!}}{e^{ - Cx}}} .
\end{equation}

Thus, the ER can be written as
\begin{equation}\label{Appendic D ergodic rate before integral}
\begin{aligned}
{R_{m,K}} = \frac{1}{{\ln \left( 2 \right)}}\sum\limits_{i = 0}^{L - 1} {\frac{{{C^i}}}{{i!}}} \int\limits_0^\infty  {\frac{{{x^i}{e^{ - Cx}}}}{{1 + x}}} dx.
\end{aligned}
\end{equation}

Hence, the tractable approximated results can be derived as
\begin{equation}\label{appendix D final result}
\begin{aligned}
{R_{m,K}} = \frac{1}{{\ln \left( 2 \right)}}\sum\limits_{i = 0}^{L - 1} {\frac{{{C^i}}}{{i!}}} \left( {\exp (C)Ei( - C) + \sum\limits_{a = 1}^i {{{( - 1)}^{a - 1}}(a - 1)!{{C}^a}} } \right).
\end{aligned}
\end{equation}
Hence, the ER of user $K$ in cluster $M$ is obtained in~\eqref{asympto W-th erogodic rate in corollary4}, and the proof is complete.

\bibliographystyle{IEEEtran}
\bibliography{IEEEabrv,NOMA_RIS_IC_arxiv}

\end{document}